\documentstyle[multicol,aps,prl,twocolumn,psfig]{revtex}
\newif\iffigs
\figstrue
\iffigs
\fi
\def\drawing #1 #2 #3 {
\begin{center}
\setlength{\unitlength}{1mm}
\begin{picture}(#1,#2)(0,0)
\put(0,0){\framebox(#1,#2){#3}}
\end{picture}
\end{center} }

\begin{document}


\twocolumn[\hsize\textwidth\columnwidth\hsize\csname@twocolumnfalse\endcsname

\title{Singularities and the distribution of density in the
Burgers/adhesion model} \author{U. Frisch$^{\rm a}$,  J. Bec$^{\rm a}$,
B. Villone$^{\rm b}$} \address{$^{\rm a}$
CNRS UMR 6529, Observatoire de la C\^ote d'Azur, BP 4229, 06304 Nice
Cedex 4, France \\$^{\rm b}$  CNR - Istituto di Cosmogeofisica, 10133
Torino, Italy} \draft
\date{\today} \maketitle

\centerline{{\it Physica} D in press}
\begin{abstract}
We are interested in the tail behavior of the pdf of mass density
within the one and $d$-dimensional Burgers/adhesion model used, e.g.,
to model the formation of large-scale structures in the Universe after
baryon-photon decoupling. We show that large densities are localized
near ``kurtoparabolic'' singularities residing on space-time manifolds
of codimension two ($d\le 2$) or higher ($d\ge3$). For smooth initial
conditions, such singularities are obtained from the convex hull of
the Lagrangian potential (the initial velocity potential minus a
parabolic term). The singularities contribute {\em \hbox{universal}
power-law  tails\/}
to the density pdf when the initial conditions are random. In one
dimension the singularities are preshocks (nascent shocks), whereas in
two and three dimensions they persist in time and correspond to boundaries
of shocks; in all cases the corresponding density pdf has the exponent $-7/2$,
originally proposed by E, Khanin, Mazel and Sinai (1997 Phys.\ Rev.\
Lett.\ 78, 1904) for the pdf of velocity gradients in one-dimensional
forced Burgers turbulence.  We also briefly consider models permitting
particle crossings and thus multi-stream solutions, such as the
Zel'dovich approximation and the (Jeans)--Vlasov--Poisson equation
with single-stream initial data: they have singularities of
codimension one, yielding power-law tails with exponent $-3$.

\end{abstract}
\vspace*{5mm}
]
\def\la{\left\langle}
\def\ra{\right\rangle}
\def\rset{{\rm I\kern -0.2em R}}
\def\un{\hbox{{1\kern -0.25em\raise 0.4ex\hbox{{\scriptsize $|$}}}}}
\def\zset{{\bf Z}}
\def\cset{eq_burg_force\hbox{{C\kern -0.55em\raise 0.5ex\hbox{{\tiny $|$}}}}}
\def\nset{\hbox{{I\kern -0.18em N}}}
\def\la{\left\langle}
\def\ra{\right\rangle}
\def\L{{\cal L}}
\def\v{{\bf v}}
\def\u{{\bf u}}
\def\x{{\bf x}}
\def\q{{\bf q}}
\def\k{{\bf k}}
\def\b{{\bf b}}
\def\rhoe{\rho ^{\rm (E)}}
\def\eql{\,{\stackrel{\rm law}{=}}\,}

{\em \noindent su quell' immenso baratro di stelle\\
sopra quei gruppi, sopra quegli ammassi,\\
quel semin\'{\i}o, quel balen\'{\i}o di stelle}\\
~\\
\noindent Giovanni Pascoli, from {\em La Vertigine} \cite{poem}


\section{Introduction}
\label{s:introduction}

In 1970 Zel'dovich introduced a simple model for explaining features of
the nonlinear formation of large-scale structures in the Universe
\cite{Zeld}.  Just after the baryon-photon decoupling in the early
Universe, there may have been a rarefied medium formed by
collisionless dustlike particles without pressure, interacting only
via Newtonian gravity \cite{peebles}.  The appropriate mathematical
description, the  equation for a self-gravitating gas in an
expanding three-dimensional universe, has so far been studied mostly
by numerical simulations.  The Zel'dovich approximation, to which we
shall return in Section~\ref{s:conclusion}, is far simpler and involves
basically particles moving in straight lines, just as rays in
geometrical optics (see also Refs.~\cite{SZ,coles-lucchin}).  As a
consequence caustics are formed, the simplest of which are {\em
pancakes},  near which the mass density is very
large. Zel'dovich, Arnold and their collaborators were mostly
interested in the nature of the singularities resulting from the model
and classified them using catastrophe theory and Lagrangian
singularity theory \cite{SZ,Arnold,Arnoldcpam,Arnoldetal}. Kofman {\it et al}.\
\cite{kofman} studied the probabilistic aspects and determined the
probability density function (pdf) $p(\rho)$ of the density of
matter. For Gaussian initial fluctuations (at decoupling) they
obtained a $\rho ^{-3}$ law for the tail at large densities. They also
gave a simple heuristic argument relating the $\rho ^{-3}$ law to the
divergence with exponent $-1/2$ near the pancakes.

One difficulty with the Zel'dovich approximation is that the pancake
structures, which are formed after the first particle crossing occurs,
rapidly smear out (see Section~\ref{s:dd}), whereas in reality massive
pancake-like structures are found to be quite long-lived.  The
gravitational dynamics of pancakes is indeed incorrectly captured
within the Zel'dovich approximation (see,
e.g., Ref.~\cite{shandarin}). This has led to the introduction of the
{\em adhesion model\/} of Gurbatov and Saichev \cite{gurb,libro} in
which particle, upon crossing, stick together (adhere). The adhesion
model is just the multi-dimensional Burgers equation, taken in the
inviscid limit $\nu\to 0$
\begin{eqnarray}
&\partial_t{\bf v}+\left({\bf v}\cdot \nabla\right) {\bf v} = 
\nu\nabla^2{\bf v}\label{burg}\\
&\v = -\nabla \psi,\label{defpot}\\
&\partial_t\rhoe+\nabla\cdot\left(\rhoe{\bf v}\right)=0.
\label{dens}
\end{eqnarray}
Here, $\v=(u,v,w)$ is the velocity, $\psi$ the (velocity) potential and
$\rhoe$ the Eulerian (mass) density (the initial Lagrangian density,
which is quasi-uniform in the cosmological problem, is denoted
$\rho_0$).  As is well known, the Burgers equation in the limit
$\nu\to 0$ produces shocks along surfaces (in three dimensions) on
which the density is infinite and across which the velocity is
discontinuous.

The question we intend to address is the behavior at large $\rho$'s of
the pdf of the density when using the one- and multi-dimensional
Burgers/adhesion model with random and smooth initial conditions,
having, e.g., Gaussian statistics with an exponentially decreasing
spectrum.  (The case of non-smooth (e.g. Brownian) initial conditions
has been considered in
Refs.~\cite{SheAurellFrisch,Sinai92,VergassolaDubrulleFrischNoullez}.)
In one dimensional decaying Burgers turbulence, the density and the
Eulerian velocity gradient have a simple relation (cf.
(\ref{gurbrel})). Hence, the density pdf is deducible from the pdf of
the velocity gradient. It was shown in Ref.~\cite{BecFrisch} that the
latter has a tail at large negative values which is a power law with
exponent $-7/2$. Actually, this law has been first proposed in
Ref.~\cite{EKMS97} for {\em randomly forced\/} Burgers turbulence. For
the forced case, the functional form of the pdf has been the subject
of considerable controversy and the question is not yet completely
resolved (see
Refs.~\cite{EKMS97,Boldyrev,BouchaudMezardParisi,EKMS98,EVandenE1,EVandenE2,tabar99,Gotoh,GotohKraichnan,Kraichnan,Polyakov,BecFrischKhanin}).
In particular E and  Vanden Eijnden \cite{EVandenE1,EVandenE2} developed a
probabilistic formalism that copes with the
delicate problems arising in the limit of vanishing viscosity
when shocks are present, proved that $\alpha < -3$ and 
made a good case for $\alpha=-7/2$.

The paper is organized as follows. In Section~\ref{s:rappels} we
present general background material about the multi-dimensional
Burgers equation and show why, paradoxically, shocks are not generally
responsible for large (but finite) densities, which generically arise
from other types of singularities. In Sections~\ref{s:oned} and
\ref{s:dd} we present the one-dimensional and the multi-dimensional
case, respectively. In the final Section~\ref{s:conclusion} we compare
the predictions of various models used in the cosmological literature:
the Burgers/adhesion model, the Zel'dovich model and the
(Jeans)--Vlasov--Poisson model.

\section{The solution of the Burgers equation and its geometrical construction}
\label{s:rappels}

We shall here be exclusively interested in the solution to the
$d$-dimensional Burgers equation (\ref{burg})-(\ref{dens}), in the
limit $\nu \to 0$, with given initial data $\v_0(\x)\equiv \v(\x,0)
=-\nabla \psi_0(\x)$ and $\rho_0\equiv\rhoe(\x,0)$.

We begin with the  velocity. The  Hopf \cite{Hopf} and Cole
\cite{Cole} transformation allows an
explicit integral representation of the solution for $\nu>0$. By  
steepest descent, a well known ``maximum
representation'' is obtained in the limit $\nu \to 0$ \cite{lax,oleinik}
\begin{equation}
\psi(\x,t) = \max_\q \left[\psi_0(\q) -{(\x-\q)^2\over 2t} \right].
\label{minimum}
\end{equation}
It is easily seen that the point $\q$ at which the maximum is
achieved is the Lagrangian point associated to the (Eulerian)
point $\x$ at time $t$. Indeed, by setting the gradient of the
r.h.s.\ of (\ref{minimum}) to zero and using (\ref{defpot}), we obtain
\begin{equation}
\x = \q +t\v_0(\q).
\label{tvezero}
\end{equation}
In other words, $\x$ is the position at time $t$ of the fluid particle
starting at $\q$ and retaining its initial velocity $\v_0(\q)$.

The problem is that (\ref{tvezero}) is valid only for {\em regular\/}
Lagrangian points, that is points which have not been captured by a
shock by time $t$. There is another 
construction of the solution which brings out the geometrical nature
of the problem with shocks. Let us define the {\em Lagrangian potential}
\begin{equation}
\varphi(\q,t)\equiv -{|\q|^2\over2}+t\psi_0(\q).
\label{lagpotdd}
\end{equation}
It follows from (\ref{minimum}) that
\begin{equation}
t\psi(\x,t)+{|\x|^2\over2}
=\max_\q\left[\varphi(\q,t)+\x\cdot\q\right]. 
\label{hlegendredd}
\end{equation}
The r.h.s.\ of (\ref{hlegendredd}) is seen to be the Legendre transform
of the Lagrangian potential \cite{Arnoldmeca}.

An important property of the Legendre transformation is that {\em the
r.h.s.\ of (\ref{hlegendredd}) is unaffected if we replace the
Lagrangian potential $\varphi(\q,t)$ by its convex hull
$\varphi_c(\q,t)$.} The Eulerian singularities of the
solution are determined by the structure of the convex hull.

Since we shall make extensive use of convex hulls in this paper, let
us define the matter precisely. Let $g(\q)$ be a real function of $\q$
defined over a convex domain ${\cal D}$ of $\rset ^d$ (for example,
the whole space). We say that $g(\q)$ is convex if, for any $\q\in
{\cal D}$,  $\q'\in {\cal D}$ and $0\le \theta\le 1$, we have
\begin{equation}
g\left(\theta \q'+(1-\theta)\q\right)\ge \theta g(\q')+(1-\theta)g(\q).
\label{defconvexg}
\end{equation}
Let $\varphi(\q)$ be an arbitrary function. We define its convex hull 
$\varphi_c(\q)$ as
\begin{equation}
\varphi_c(\q) \equiv \min g(\q),
\label{deffc}
\end{equation}
the minimum being taken over all functions $g(\cdot)\ge \varphi(\cdot)$
which are convex.

\begin{figure}
\iffigs 
\centerline{\psfig{file=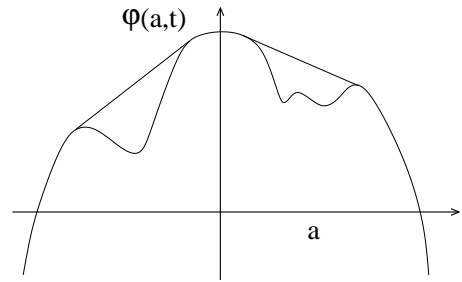,width=6cm}}
\else\drawing 65 10 {Lagrangian potential 1D}
\fi
\vspace{2mm}
\caption{Lagrangian potential and its convex hull in one
dimension. The graph of the convex hull contains regular parts of the
Lagrangian potential and segments touching the original graph at two
points, lying over shock intervals.}
\label{f:convex-hull-1d}
\end{figure}

\begin{figure}
\iffigs 
\centerline{\psfig{file=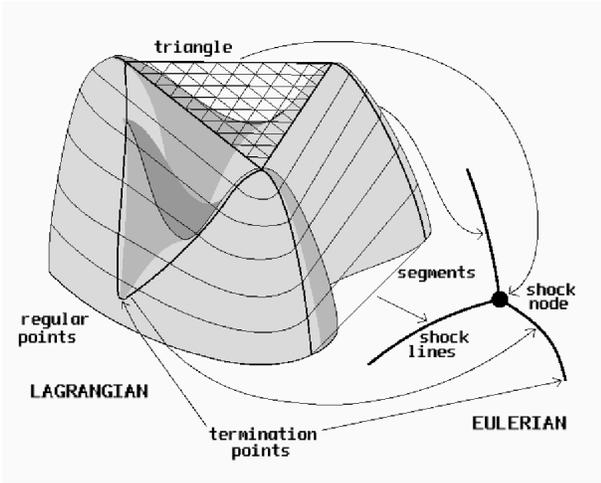,width=8cm}}
\else\drawing 65 10 {convex-hull construction 2D}
\fi
\vspace{2mm}
\caption{The convex-hull construction in two dimensions (adapted
from Ref.~\protect\cite{VergassolaDubrulleFrischNoullez}); the graph of the
convex hull contains regular parts of the Lagrangian potential
(corresponding to regular points in the Eulerian space), pieces of
ruled surfaces (corresponding to shock lines), triangles (corresponding
to shock nodes) and ``kurtoparabolic points'' (corresponding to
termination points of shock lines).}
\label{f:convex-hull-2d}
\end{figure}

In other words, the graph of $\varphi_c(\q)$ is obtained by tightly
pulling a string (in one dimension) or an elastic sheet (in two
dimensions) around the graph of $\varphi(\q)$ (see
Figs.~\ref{f:convex-hull-1d} and \ref{f:convex-hull-2d}). 

In one dimension the graph of the convex hull contains (i) parts of
the graph of $\varphi(\q)$ and (ii) segments touching the original
graph at two points.  The former correspond to regular (Lagrangian)
points and the latter to points having fallen into a shock. In two
dimensions the convex hull consists generically of four kind of
objects: (i) parts of the original graph, (ii) pieces of ruled
surfaces, (iii) ``kurtoparabolic points'', to which we shall come
back, and (iv) triangles (see Fig.~\ref{f:convex-hull-2d}). The
associated Eulerian objects are, respectively, (i) regular points,
(ii) shock lines, (iii) end points of shocks and (iv) shock nodes.
For the three-dimensional case, see Ref.~\cite{boga} and references 
therein.

A more complete description of singularities is obtained by considering
the metamorphoses  of singularities as
time elapses. A complete classification in two and three dimensions may 
be found in the appendix (supplement 2) by V.I.~Arnold, Yu.M.~Baryshnikov
and I.A.~Bogayevski of Ref.~\cite{libro}. 

Let us just observe at this stage that, when the initial potential
$\psi_0(\q)$ is a sufficiently smooth function of $\q$ (as we
assume in this paper), the Lagrangian potential given by
(\ref{lagpotdd}) is necessarily convex when $t$ is sufficiently small.
Indeed, the Hessian (determinant of the Hessian matrix of second space
derivatives)
\begin{equation}
H(\q,t) \equiv \det \left({\partial ^2 \varphi(\q,t)\over \partial
q_i\partial q_j} \right)= t^d \det \left(-{1\over t}\delta_{ij}+
{\partial ^2
\psi_0(\q)\over  \partial q_i\partial q_j} \right)
\label{hessian}
\end{equation}
remains very close to its initial value $(-1)^d$ for short times.
As long as the Hessian does not change sign, the convexity is the same as for
the initial paraboloid $\varphi(\q,0)= -|\q|^2/2$.

We also introduce two Lagrangian maps from $\q$ to $\x$. The 
{\it naive Lagrangian map\/} $L_t$ is just given by (\ref{tvezero}). The
(proper) {\it   Lagrangian map\/} is
\begin{equation}
\L_t:\q\mapsto -\nabla_\q\varphi_c(\q,t)=\x(\q,t).
\label{lagmapdd}
\end{equation} 
At regular points, where $\varphi(\q,t)$ and $\varphi_c(\q,t)$
coincide, so do the two Lagrangian maps. If, however, $\q$ is not a
regular point, the Lagrangian map transforms it into the appropriate
Eulerian shock location, whose determination requires the knowledge of
the convex hull, that is, a global geometrical construction. The map
$\q\mapsto \x=\L_t \q$ is invertible only at Eulerian points which are
not on shocks (otherwise there is more than one Lagrangian point which
is mapped into $\x$).  The Jacobian of the Lagrangian map at regular points
is defined as
\begin{equation}
J(\q,t)\equiv \det \left(\partial x_i\over \partial q_j \right).
\label{defj}
\end{equation}
It follows from (\ref{hessian}) and (\ref{lagmapdd}) that it
is just $(-1)^d$ times the Hessian $H(\q,t)$ of the Lagrangian
potential.  

We turn now to the determination of the (Eulerian) density
$\rhoe(\x,t)$.
Mass conservation implies that, if $\x$ is not on a shock,  
\begin{equation}
\rhoe(\x,t) = {\rho_0\over J(\q,t)}, \qquad \q = \L_t ^{-1}\x.
\label{conserv}
\end{equation}
If $\x$ is on a shock, the density is of course infinite. This does
not, however, imply that {\em large but finite\/} densities are
generally obtained near shocks. Indeed, for $\rhoe(\x,t)$ to be large,
the Jacobian $J(\L_t ^{-1}\x,t)$ and thus the Hessian $H(\L_t
^{-1}\x,t)$ must be small. At any given time $t$, this happens only
near the $(d-1)$-dimensional manifold of vanishing Hessian.
Arbitrarily close to such a ``parabolic'' point there are generically
hyperbolic points where the surface defined by $\varphi(\q)$ crosses
its tangent (hyper)plane and which, therefore, do not belong to its
convex hull. As we shall see in Sections~\ref{s:oned} and \ref{s:dd}
there can be ``kurtoparabolic points'' with vanishing Hessian which
are at the boundary of regular regions. (In Greek,
$\kappa\upsilon\rho\tau o\sigma$ means ``convex''; hence the proposed
name. In Ref.~\cite{boga} they are called $A_3$-points.)  In one dimension kurtoparabolic points correspond to preshocks
\cite{fofr83} and exist only at discrete times; in two and three
dimensions they generally persist for a finite time and are
associated, in the Eulerian space, to boundaries of shocks
(termination points in two dimensions and edges in three dimensions).
Large densities are obtained exclusively in the neighborhood of
kurtoparabolic points.

\section{The one-dimensional case}
\label{s:oned}

\subsection{Preshocks}
\label{s:preshocks}

In one dimension, following fluid dynamical tradition, we denote the
Lagrangian coordinate by $a$ and the velocity by $u$. We denote by
$u_0(a)=-d\psi_0(a)/da$ the initial velocity, which is assumed to be
random and sufficiently smooth. It is convenient, but not essential,
to assume periodic boundary conditions, homogeneity and a vanishing
mean velocity. We denote by $\rho_0$ the initial background density,
taken deterministic and uniform. At regular points (outside of shocks)
the Eulerian velocity and density are given implicitly by
\begin{eqnarray}
&&u(x,t)=u_0(a), \qquad \rhoe(x,t)= {\rho_0\over \partial_a x}, 
\label{urho1d}\\
&&x= a +tu_0(a).
\label{eulerlag1d}
\end{eqnarray}
Since $\partial_x u =\left(\partial_a
u_0\right)\left(\partial_a x\right)^{-1}$, it follows that the Eulerian
density can be written in terms of the Eulerian velocity gradient
$\partial_xu$ \cite{libro}
\begin{equation}
\rhoe(x,t)= \rho_0\left(1-t\partial_xu(x,t)\right).
\label{gurbrel}
\end{equation}
In Ref.~\cite{BecFrisch} it was shown that the pdf of the Eulerian
velocity gradient has a -7/2 power law at large negative values.
Hence,  the pdf $p(\rho)$ of $\rhoe$ has also a  -7/2
power law, but  at large positive values. The proof given in
Ref.~\cite{BecFrisch} was rather detailed. Here, we give a simplified
derivation, adapted to the case of the density. Furthermore, we shall
work mostly with the potential and normal forms near singularities, to
prepare the ground for the multi-dimensional case.

From (\ref{urho1d}) and (\ref{eulerlag1d}) we have, at regular points,
\begin{eqnarray}
\rhoe(L_ta,t)={\rho_0\over 1-td^2\psi_0(a)/da ^2}.
\label{rhoepsi}
\end{eqnarray}
Large values of $\rhoe$ are thus obtained in the neighborhood of
Lagrangian points with vanishing Jacobian, where $d^2\psi_0(a)/da
^2=1/t$. Once mature shocks have formed, the Lagrangian points with
vanishing Jacobian are {\em inside shock intervals and thus not
  regular.} The only kurtoparabolic points (points with vanishing
Jacobian at the boundary of regular regions) are obtained at
preshocks, that is when a new shock is just born at some time $t_*$.
Preshocks, play a central role in the -7/2 law of E, Khanin, Mazel and
Sinai  for the forced case \cite{EKMS97}. Such points, denoted by
$a_*$, are local negative minima of the initial velocity gradient,
characterized by the following relations
\begin{equation}
{d^2\psi_0\over da ^2}(a_*)={1\over t_*},\quad {d^3\psi_0\over da
^3}(a_*)=0,
\quad {d^4\psi_0\over da ^4}(a_*)<0.
\label{3conditions}
\end{equation}
There is however an additional global regularity condition that the
preshock point  $a_*$ has not been captured before $t_*$ by a mature
shock. This may be written  in terms of the convex hull $\varphi_c$ of
the  Lagrangian potential $\varphi$, as
\begin{equation}
\varphi(a_*,t_*)=-a_* ^2/2 +t\psi_0(a_*)=\varphi_c(a_*,t_*).
\label{convexastar}
\end{equation}
As shown in \cite{BecFrisch}, this global condition affects only
constants and not the scaling properties of $p(\rho)$ at large
$\rho$'s.

\begin{figure}
\iffigs 
\centerline{\psfig{file=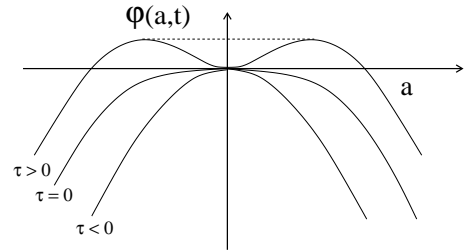,width=6cm}}
\else\drawing 65 10 {Normal form prechoc 1D}
\fi
\vspace{2mm}
\caption{Normal form of the Lagrangian potential in the neighborhood
of a preshock in one dimension. At the time of the preshock
($\tau=(t-t_*)/t_*= 0$), the Lagrangian potential changes from a
single extremum to three extrema and develops a non-trivial convex
hull (shown as a dashed line).}
\label{f:lagpot1d}
\end{figure}

We can now Taylor expand the Lagrangian potential and the Lagrangian
map near the space-time location $(a_*,t_*)$. By adding a suitable
constant to the initial potential, by performing a suitable
translation and also a Galilean transformation canceling the
initial velocity at $a_*$, we may assume that
$a_*=\psi_0(a_*)=d\psi_0(a_*)/da =0$. We then obtain, to the relevant
leading order, the following ``preshock normal forms''
\begin{eqnarray}
\varphi(a,t)&\simeq& {\tau a ^2\over2} + \zeta a^4,
\label{devpotlag}\\
x(a,t)&\simeq& -\tau a-4\zeta a^3, \label{taylorx}\\
J(a,t)&\simeq&-\tau  -12\zeta a^2,
\label{devjac}
\end{eqnarray}
where
\begin{equation}
\tau \equiv {t-t_*\over t_*},\qquad \zeta\equiv {t_*\over24}{d^4\psi_0\over da
^4}(0)<0.
\label{deftauzeta}
\end{equation}
The Lagrangian potential, together with its convex hull, are shown in
Fig.~\ref{f:lagpot1d}. 
Note that at $t=t_*$ there is a degenerate
maximum with quartic behavior and that, immediately after $t_*$, for $\tau>0$,
convexity is lost and  a shock interval is born. Given the symmetry,
resulting from our choice of coordinates, the convex hull contains
a horizontal segment extending between the two maxima $a_\pm =
\pm (-\tau/(4\zeta))^{1/2}$. Note that for $\tau>0$ the Jacobian
vanishes at two locations $\pm (-\tau/(12\zeta))^{1/2}$ which are
within the shock interval and are therefore irrelevant as far as
the Burgers/adhesion model is concerned (although they become relevant
when particle crossing is permitted; see Section~\ref{s:conclusion}).

From (\ref{devjac}) we see that the density $\rho_0 /J$ has a $a
^{-2}$ singularity in Lagrangian coordinates at $t=t_*$
($\tau=0$). Since, by (\ref{taylorx}), the relation between $a$ and
$x$ is cubic at $\tau=0$, the density $\rhoe(x,t_*)\propto |x|^{-2/3}$
which is unbounded. For any $t\neq t_*$ the density remains bounded,
except at the shock location. For $\tau<0$, this follows immediately
from (\ref{devjac}), which implies $\rhoe \le \rho_0/|\tau|$. For
$\tau>0$, the exclusion of the shock interval requires  $|a|>a_+$.
Hence,  $\rhoe \le \rho_0/(2\tau)$. It is clear that large densities
are obtained only in the immediate neighborhood of the preshock.
More precisely, it follows from (\ref{taylorx}) and (\ref{devjac})
that $\rhoe>\rho$ requires simultaneously
\begin{equation}
|\tau|<{\rho_0\over\rho}\,\,\,\,{\rm
and}\,\,\,\,|x|<(-12\zeta)^{-1/2}\left({\rho_0\over\rho}\right)^{3/2},
\label{bornes}
\end{equation}
which become very small intervals around the spatio-temporal location
of preshocks when $\rho$ is large.

\subsection{The -7/2 law in one dimension}
\label{s:7halvlaw}

So far, we have looked at the question of large densities from a
deterministic point of view. We turn to the probability to have a
large density at a {\em given\/} Eulerian point $x$ and a given time
$t$. We shall calculate the cumulative distribution
\begin{equation}
P^>(\rho;x,t)\equiv {\rm Prob}\,\left\{\rhoe(x,t)>\rho\right\},
\label{defcum}
\end{equation}
from which we obtain the pdf $\,p(\rho;x,t)=-\partial_\rho
P^>(\rho;x,t)$.  In the random case each preshock has a random
Eulerian location $x_*$, occurs at a random time $t_*$ and has a
random $\zeta<0$ coefficient (there is also a random velocity $u_*$ of
the preshock but this is easily seen to be irrelevant for our
purposes). Only those realizations such that $x_*$ and $t_*$ are
sufficiently close to $x$ and $t$ will contribute large densities.
Denoting by $p_3(x_*,t_*,\zeta)$ the joint pdf of the three arguments,
which is understood to vanish unless $\zeta<0$, we have
\begin{equation}
P^>(\rho;x,t) =\int_{\rhoe(x,t)>\rho}p_3(x_*,t_*,\zeta)\,dx_*\,dt_*\,d\zeta.
\label{psupdep}
\end{equation}
(If homogeneity is assumed $p_3$ does not depend on $x_*$; the case of
homogeneity extending over the whole space can be obtained by letting
the spatial period $L\to \infty$; we must then also replace $p_3$ by
$n(t_*\zeta)/L$ where $n$ is a number of preshocks per unit length
rather than a probability; similarly, $P^>$ is then a probability per
unit length.) Because of the very sharp localization near preshocks
implied by (\ref{bornes}), for large $\rho$'s, we may replace
$p_3(x_*,t_*,\zeta)$ by $p_3(x,t,\zeta)$. Using then, in a suitable
frame, the normal forms (\ref{devpotlag})-(\ref{devjac}) we can
rewrite (\ref{psupdep}) as an integral over local Lagrangian variables
$a$ and $\tau$ and obtain
\begin{equation}
P^>(\rho;x,t)\simeq \int_D t\,(-\tau -12\zeta a ^2)\,p_3(x,t,\zeta)\,da\, d\tau\, d\zeta.
\label{escroc}
\end{equation}
Here, the domain $D$ is the set of $(a,\tau,\zeta)$ such that 
\begin{equation}
{\tau\over -4\zeta}<a ^2< {1\over -12\zeta}\left({\rho_0\over\rho} +\tau\right)
\label{jacobetc}
\end{equation}
The right part of (\ref{jacobetc}) expresses that the density exceeds
the value $\rho$, while the left part (which is trivial when $\tau<0$
since $\zeta<0$) excludes the shock interval $]a_-,a_+[$. In
(\ref{escroc}) the factor $-\tau -12\zeta a ^2$ is a Jacobian stemming
from the change to Lagrangian space variables and the factor $t$ stems
from the change of temporal variables. The integration over $a$ and
$\tau$ can be carried out explicitly, yielding 
\begin{eqnarray}
P^>(\rho;x,t)&\simeq& C(x,t)\left({\rho_0\over\rho}\right)^{5/2},
\label{cayest}\\
C(x,t)& \equiv &
At\int_{-\infty}^0
|\zeta|^{-1/2} p_3(x,t,\zeta)\,d\zeta,
\label{cayestC}
\end{eqnarray}
where $A$ is a positive numerical constant.  Thus, for any $x$ and $t$, the
cumulative probability of the density follows a $\rho^{-5/2}$ law. Hence,
$p(\rho;x,t)\propto \rho^{-7/2}$, as $\rho\to\infty$, which establishes the
$-7/2$ law for the pdf. Note that, contrary to the derivation in
Ref.~\cite{BecFrisch}, we did not use homogeneity.  With this additional
assumption, $p_3$ and thus $C(x,t)$ become independent of $x$. 

Taking into account higher-order singularities does  not influence
this result. Consier, e.g., quintic-root
preshocks arising from degenerate inflection points in the initial velocity, at
which the second, third and fourth space derivatives all vanish. In one
dimension such singularities are not generic in the deterministic case but
could nevertheless contribute in the random case, as happens in Berry's 
``battle of catastrophes'' \cite{berry}. The exact vanishing of two more 
derivatives has probability zero but there is a finite probability that 
the third and fourth velocity derivatives have values small enough 
to give the preshock an approximately quintic-root structure. We found that
such events contribute only subdominant corrections to the -7/2 law.

\section{Two dimensions and beyond}
\label{s:dd}

\subsection{Preshocks in two dimensions}
\label{s:pre2d}

In two dimensions we use the notation $\q =(a,b)$ and $\x=(x,y)$ for
Lagrangian and Eulerian coordinates. It follows from (\ref{hessian})
that the first singularity happens at the time $t_*$ which is the
inverse of the largest positive eigenvalue of the Hessian matrix of
the initial potential $\psi_0(a,b)$. No generality is lost by making
the following assumptions: (i) the maximum is achieved at the origin,
(ii) the potential and its gradient vanish at the origin, (iii) the
maximum eigenvalue is equal to one ($t_*=1$) and (iv) the
eigendirections of the matrix of second derivatives are the $a$-axis
for the eigenvalue 1 and the $b$-axis for the other eigenvalue
$1-\mu $ with $\mu>0$. Using this, we can Taylor expand the initial
potential to the relevant (fourth) order:
\begin{eqnarray}
\psi_0(a,b) &\simeq& {a ^2\over2} +(1-\mu) {b ^2\over2}+\alpha a ^3 +\beta a
^2b +\gamma a b ^2 +\delta b ^3\nonumber\\
&&\,\,\,+ \zeta a ^4+\eta a ^3 b+\theta a ^2 b ^2 +\kappa a b ^3 +\rho b ^4.
\label{taylorpsi2d}
\end{eqnarray}
Expressing that the matrix of second derivatives has its largest
eigenvalue at $a=b=0$, we find that $\alpha=\beta =0$ and that
the quadratic form 
\begin{equation}
Q(a,b)\equiv 12\mu\zeta a ^2+6\mu\eta ab+(2\mu\theta+ 4\gamma ^2)b^2
\label{laforme}
\end{equation}
must be definite negative, thereby putting certain
restrictions on $\mu$,  $\zeta$, $\theta$ and $\gamma$ which will
henceforth be assumed. 

\begin{figure}
\iffigs 
\centerline{\psfig{file=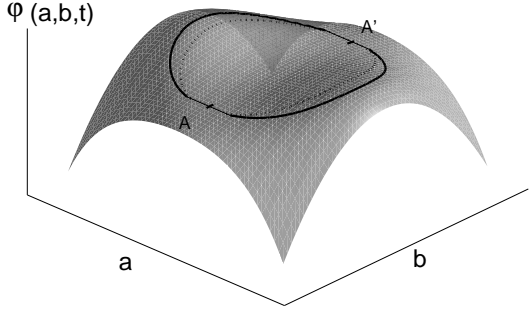,width=7cm}}
\else\drawing 65 10 {Normal form preshock 2D}
\fi
\vspace{2mm}
\caption{The Lagrangian potential just after a preshock
in two dimensions. The continuous line is the separatrix between the
regular part and the ruled surface of the convex hull. The dotted line
corresponds to the vanishing of Jacobian of the Lagrangian map. They
touch at two ``kurtoparabolic'' points A and A'.}
\label{f:deuxbosses}
\end{figure}

We can now write the corresponding normal form of the Lagrangian
potential $t\psi_0(a,b)-(a ^2+b^2)/2$, at time $t=1+\tau$
for small  $\tau$,
which includes all the relevant terms
\begin{equation}
\varphi(a,b,t) \simeq \tau {a ^2\over2} -\mu {b ^2\over2} 
+\gamma a b ^2 + \zeta a ^4+\eta a ^3 b+\theta a ^2 b ^2 .
\label{normalpotlag2d}
\end{equation}
For $\tau<0$ the surface defined by the Lagrangian potential has a
single maximum at $a=b=0$. At $\tau =0$ this maximum still exists
but is quartically degenerated in the $a$-direction. For $\tau >0$,
the origin turns into a saddle and two new maxima appear (for 
values of $a=O\left(\tau ^{1/2}\right)$ and of $b=O\left(\tau
^{3/2}\right)$). The general aspect of the surface is shown in 
Fig.~\ref{f:deuxbosses}. It is clearly not convex; hence, a new shock 
is born.

The Hessian, now equal to the Jacobian of the Lagrangian map, which
controls the density at regular points is given by
\begin{equation}
H(a,b,t)\simeq -\tau\mu  -Q(a,b),
\label{2dhessian}
\end{equation}
where $Q$ is given by (\ref{laforme}). 

Consider first the situation at $t=t_*$. The Hessian vanishes and,
hence, the density becomes infinite at the origin. This is the
singularity called $A_3$ by Arnold \cite{Arnold,Arnoldetal,Arnoldmeca},
for which the mean density in a small disk of radius $r$ around the
origin (in the the Eulerian space) is easily shown to diverge as
$\bar\rho(r)\propto r^{-2/3}$.

Shortly after $t_*$, if we (incorrectly) use $\varphi(a,b,t)$ rather than its
convex hull, we find that the density becomes infinite in Lagrangian
coordinates along a zero-Hessian curve of approximately elliptical shape. On
the surface defined by the Lagrangian potential $\varphi(a,b,t)$, the
corresponding points are parabolic (they are shown as a dotted line on
Fig.~\ref{f:deuxbosses}). When approaching such a line of parabolic points,
one has an $A_2$ singularity in the sense of Arnold, for which
$\bar\rho(r)\propto r^{-1/2}$. Actually, nearly all the points on this
line are ``hidden under the convex hull''. Constructing the convex hull of
$\varphi(a,b,t)$ is not a local operation and thus, in general, not
elementary. Let us just illustrate what can happen in the simpler case where
$\gamma=\eta=0$, which has an additional symmetry.  We then have
\begin{equation}
\varphi(a,b,t) \simeq \tau {a ^2\over2} -\mu {b ^2\over2} 
+ \zeta a ^4 +\theta a ^2 b ^2 ,
\label{normalpotsym}
\end{equation}
which is even in both $a$ and $b$. The conditions of negative
definiteness of the quadratic form $Q$ are
then $\zeta<0$ and $\theta<0$.  It follows from the symmetries and the
fact that lines of constant $a$ are parabolas that the convex hull contains
a piece of ruled surface made of segments parallel to the
$a$-axis. These segments are connecting the two maxima of sections at
constant $b$, which exist for any $b^2<-\tau/(2\theta)$.  The
horizontal projections of these end points define the {\em separatrix\/}
between regular Lagrangian points and points absorbed into the newly
created shock. It is the  ellipse 
\begin{equation}
\tau+4\zeta a ^2+2\theta b^2 =0,
\label{separatrix}
\end{equation}
obtained by requiring $\partial_a\varphi(a,b,t)=0$.
The associated points on the surface are shown as a continuous line
on Fig.~\ref{f:deuxbosses}. The corresponding Eulerian structure is
easily seen to be an embryonic shock line, parallel to the $y$-axis,
with a length $O\left(\tau ^{1/2}\right)$ and a velocity jump also
$O\left(\tau ^{1/2}\right)$, except near its end points, where it vanishes.

It is now easily checked that the  separatrix ellipse and the zero-Hessian
ellipse are tangent at the points $a=0,\,\,b=
\pm \left(-\tau/(2\theta)\right)^{1/2}$, denoted A and A' on  
Fig.~\ref{f:deuxbosses}. These points of vanishing Hessian, which
belong to the edge of the regular region, are the only kurtoparabolic points
in the sense of Section~\ref{s:rappels}. Arbitrarily large densities
are obtained in their neighborhood. Contrary to the one-dimensional
case, the condition $\rhoe>\rho$, for large $\rho$, does not put an
upper bound, similar to (\ref{bornes}), on the time $\tau$ elapsed since $t_*$.
Actually, in two (and more) dimensions, kurtoparabolic points persist
generically for at least a finite time, irrespective of the presence of the
additional symmetry assumed in the simple example given above.

\subsection{Kurtoparabolic points in two dimensions}
\label{s:kuku}

We recall our definition of a kurtoparabolic point A as a point (i)
where the Hessian of the Lagrangian potential $\varphi(a,b,t)$
vanishes and (ii) which belongs to the boundary of the regular part of
the convex hull $\varphi_c(a,b,t)$. This requires two local
constraints: that A be parabolic and that the surface defined by
$\varphi(a,b,t)$ should not cross the tangent plane at A. It also
requires a global constraint, namely that A should not be situated
below a piece of the convex hull, which would correspond to A having
been absorbed by a mature shock before the current time $t$. The
latter condition is automatically satisfied at the birth of the first
singularity and will then persist for at least a finite time. The
former can be expressed purely in terms of the local properties of the
Lagrangian potential. For this, let us find the normal form associated
to a kurtoparabolic point at an arbitrary time $t$ (not necessarily
close to $t_*$). In what follows the time is purely a parameter which
will not be written. The vanishing of the Hessian requires the
vanishing of at least one eigenvalue of the Hessian matrix of second
derivatives.  We assume, again without loss of generality, that A is
the origin, that the potential and its gradient vanish at A, that the
vanishing eigenvalue of the Hessian matrix corresponds to the $a$-axis
and that the other eigenvalue is $-\mu <0$. We now write the local
Taylor expansion of the Lagrangian potential. Clearly, we need to
include terms up to fourth order in $a$, but we shall see that the
relevant order in $b$ is two (because $b=O(a ^2)$). Hence, we can
write
\begin{equation}
\varphi(a,b) \simeq -\mu {b ^2\over2}+\alpha a ^3 +\beta a
^2b + \zeta a ^4.
\label{taylorkuku}
\end{equation}
We now require that, at least locally (i.e.\ for small $a$ and $b$),
the surface defined by $\varphi(a,b)$ should be below its tangent plane
at the origin. This amounts just to $\varphi(a,b)\le 0$ and is equivalent to
\begin{equation}
\alpha =0,\qquad \zeta<0, \qquad \beta ^2< -2\mu \zeta.
\label{diagennul}
\end{equation}
(Note that $\beta\neq 0$ since its vanishing would correspond to
having a preshock.)  The only ``sharp'' condition is the vanishing of
the coefficient $\alpha$ of $a ^3$. Indeed, for $\alpha \neq 0$ the surface
crosses its tangent plane at the origin.  Hence, there are two sharp
conditions for the existence of a kurtoparabolic point: the vanishing
of the Hessian and of the coefficient of $a ^3$.  Thus, we expect that
kurtoparabolic points are found on codimension two manifolds in
space-time, that is, time-dependent discrete locations which persist
for at least a finite time.

Actually, kurtoparabolic points are generally born with
the first singularity which is itself such a point, albeit one 
with a higher degree of degeneracy  (the coefficient $\beta $ also
vanishes). Generically, kurtoparabolic points disappear at large times when
only a network of shocks subsists. This is seen, for example, in
the study of flame-front cracks in Ref.~\cite{kuznetsov}. 

\begin{figure}
\iffigs 
\centerline{\psfig{file=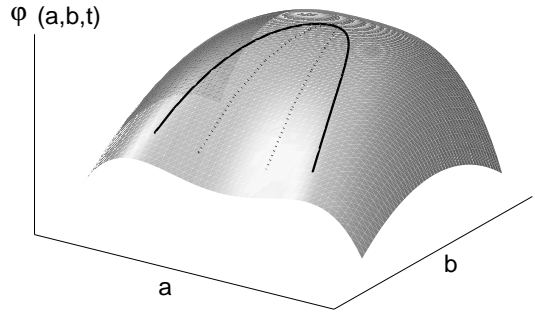,width=7cm}}
\else\drawing 65 10 {kurtoparabolic points 2D}
\fi
\vspace{2mm}
\caption{The Lagrangian potential in two dimensions, in
the neighborhood of a kurtoparabolic point. The continuous line is the
separatrix between the regular part and the ruled surface of the
convex hull. The dotted line corresponds to the vanishing of the
Jacobian of the Lagrangian map.}
\label{f:boutdechoc}
\end{figure}

The typical local aspect of a kurtoparabolic point is shown in
Fig.~\ref{f:boutdechoc}.
The $a\mapsto -a$ symmetry, which is here
generic, and the convexity of sections by planes of constant $a$
allows a straightforward construction of the convex hull of
$\varphi(a,b)$.  The convex hull contains, as in
Section~\ref{s:pre2d}, a ruled surface made of segments parallel to
the $a$-axis. The separatrix  is now a parabola of equation
\begin{equation}
b=-{2\zeta\over\beta} a ^2,
\label{kukusep}
\end{equation}
shown as a continuous line in Fig.~\ref{f:boutdechoc}. The line of
vanishing Hessian 
\begin{equation}
H(a,b)= -2\mu\beta b-4\left(3\mu\zeta +\beta ^2\right) a ^2=0
\label{hessiankuku}
\end{equation}
is also a parabola (shown as a dotted line) which
touches the former at the kurtoparabolic point. The corresponding
Eulerian structure is a shock line with an end point of the kind shown
in Fig.~\ref{f:convex-hull-2d}. High densities $\rhoe=\rho_0/H(a,b)$ are obtained
in the neighborhood of the kurtoparabolic point. Specifically,
the set of regular Lagrangian point such that $\rhoe>\rho$ is defined 
by the following inequalities
\begin{equation}
-{\beta b\over2\zeta}<a ^2<-{\rho_0/\rho+2\mu\beta
b \over 4\left(3\mu\zeta + \beta ^2\right)}.
\label{rhoesuprhobout}
\end{equation}
Using the Lagrangian map $x= -\partial_a \varphi,\,\,y=  -\partial_b
\varphi$ it may be shown that, in the Eulerian space, the mean density in 
a disk of small radius $r$ around the end of the shock line
$\bar\rho(r)\propto r^{-2/3}$.

It follows from (\ref{rhoesuprhobout}) that, when $\rho$ is large 
$a$ and $b$ are restricted to being very close to the kurtoparabolic point.

\subsection{The -7/2 law in two dimensions}
\label{s:7halv2d}

We now assume random and smooth initial conditions and proceed along
the same general lines as in Section~\ref{s:7halvlaw}; in particular, spatial
periodicity is  assumed (here and in the next section), although this is
not essential. The cumulative
probability to have $\rhoe>\rho$ is now expressed in terms
of the joint pdf of all the relevant parameters at a kurtoparabolic
point (its position $\x_*$, and the three parameters $\mu$, $\beta$ and
$\zeta$) in the normal form:
\begin{equation}
P^>(\rho;\x,t) =\int_{\rhoe(\x,t)>\rho}p_4(\x_*,\mu,\beta,\zeta;t)
\,d\x_*\,d\mu\,d\beta\,d\zeta.
\label{psupdep2d}
\end{equation}
(Additional dependences of the probability density on the velocity at
the kurtoparabolic point and on the orientation of the axis
corresponding to the vanishing eigenvalue are omitted for simplicity,
since they are irrelevant for the calculation of the density.)  An
important difference with the one-dimensional case, due to the
persistence of kurtoparabolic points, is the lack of a $t_*$ argument
in $p_4$. The probability to have a kurtoparabolic point at $\x_*$ may
still depend on $\x_*$ (unless the initial condition is
homogeneous) and will certainly depend on the time $t$. With Gaussian
initial conditions it may be shown that $p_4$ is non-vanishing for any
$t>0$.

Next, we use the very sharp localization of high-densities near the
Eulerian points which are associated to the kurtoparabolic points and
change from Eulerian to (local normal form) Lagrangian coordinates,
rewriting (\ref{psupdep2d}) as
\begin{equation}
P^>(\rho;\x,t)\simeq \int_{\cal D} H(a,b,t)\,p_4(\x,\mu,\beta,\zeta;t)\,da
\,db\,d\mu\,d\beta\,d\zeta,
\label{escroc2d}
\end{equation}
where the domain  ${\cal D}$ is the set of $(a,b,\mu,\beta,\zeta)$
such that (\ref{rhoesuprhobout}) holds and $H(a,b,t)$, given by
(\ref{2dhessian}), is the Jacobian of the Lagrangian map. In
(\ref{escroc2d})
the integration over $a$ and $b$ can be carried out explicitly. This
gives $P^>(\rho;\x,t)\propto (\rho_0/\rho)^{5/2}$ and, hence,
\begin{equation}
p(\rho;\x,t)\propto \left({\rho\over\rho_0}\right)^{-7/2},\quad \rho\to\infty,
\label{laloi}
\end{equation}
which is the two-dimensional -7/2 law. (The constant in front of the
power law, which involves integrals over $\mu$, $\beta$ and $\zeta$, is
not written.)

It is of interest to note that we have obtained exactly the same
scaling as in one dimension. The reason for this is rather
interesting. In one dimension, the dominant singularities are
preshocks which are isolated events in space-time, while in two
dimensions they are kurtoparabolic points which are persistent in
time. Nevertheless, if we compare the integrals (\ref{escroc}) and
(\ref{escroc2d}) and the conditions on the integration domains 
(\ref{jacobetc}) and (\ref{rhoesuprhobout}), we find that, in two
dimensions the spatial $b$-variable plays exactly the same role as
the temporal $\tau$-variable in one dimension. We can see this also
by examining Fig.~\ref{f:boutdechoc}: cuts at constant $b$ will have
the same shape as the curves shown in Fig.~\ref{f:lagpot1d} for
the one-dimensional case, changing from a single maximum for $b<0$,
to a quartically degenerate maximum for $b=0$ and to two symmetrical
maxima for $b>0$.

We must  stress that in two (and more) dimensions, the dominant
contribution to the tail of the density does not come from preshocks as 
in one dimension but comes
from the entire life span of kurtoparabolic points which are just 
born at preshocks. Actually, the contribution of a small
time interval straddling a preshock gives a $\rho ^{-4}$ intermediate
asymptotic law, going over to a 
$\rho ^{-7/2}$ law at very  large $\rho$.

\subsection{Higher dimensions}
\label{s:higher}

In dimensions $d>2$, kurtoparabolic points are now space-time manifolds of 
codimension two. The corresponding normal form (in
suitable coordinates) can be written, using  $a$ for the coordinate
in the direction of the vanishing eigenvalue of the Hessian matrix and
$\b =b_i$ ($i=1,\ldots, d-1$) for  the $d-1$ remaining coordinates. The
analog of (\ref{taylorkuku}) with a vanishing  $\alpha a ^3$
term is now:
\begin{equation}
\varphi(a,\b) \simeq   \zeta a ^4 +\sum_{i=1}^{d-1} \left[ -\mu_i
{b_i ^2\over2}+\beta_i a ^2b_i \right],
\label{taylorkukuhigh}
\end{equation}
where 
\begin{equation}
\zeta<0, \qquad \mu_i>0 \qquad \sum_{i=1}^{d-1}{\beta_i ^2\over
\mu_i} <-2\zeta
\label{diagennuldd}
\end{equation}
are the analogs of (\ref{diagennul}).  An easy calculation shows that
the Jacobian $J=(-1)^d H(a,\b)$ of the Lagrangian map is given by
\begin{equation}
J(a,\b)=-\left(\prod_{i=1}^{d-1}\mu_i\right)\left[ 4\left( 3\zeta+\sum
_{i=1}^{d-1}{\beta_i ^2 \over \mu_i} \right) a ^2 +2\sum _{i=1}^{d-1}
\beta_i b_i \right ].
\label{jacobiendd}
\end{equation}
Because of the $a\to -a$ symmetry and the convexity in all the other
variables,  the separatrix of the convex hull is again obtained by
setting $\partial _a \varphi(a,\b)=0$, thereby obtaining
\begin{equation}
2\zeta a ^2 +\sum _{i=1}^{d-1}\beta_ib_i=0.
\label{separatrixdd}
\end{equation}
By proceeding as in two dimensions, we can write the analog of
(\ref{escroc2d}) in which $b$, $\mu$ and $\beta$ must now be
reinterpreted as $(d-1)$-dimensional vectors. As to the integration
domain ${\cal D}$, it is now defined by $J(a,\b)<\rho_0/\rho$ and the
negativity of the l.h.s.\ of (\ref{separatrixdd}), which expresses the
belonging to the regular domain.

We now observe that $\sum _{i=1}^{d-1}\beta_ib_i$ plays here the same
role as $\beta b$ in the two-dimensional case. There is no
$\b$-dependence other than this. Hence, assuming that not all the
$\beta_i$'s vanish (otherwise we have a preshock) we can change
variables in the $\b$-space, taking one axis in the direction of $\sum
_{i=1}^{d-1}\beta_ib_i$. The integration in all the $\b$-directions
orthogonal to this direction is trivial and gives order unity
contributions. The remaining integral is just the same as in two
dimensions. Hence, in any dimension $d>2$, the
contribution stemming from such kurtoparabolic manifolds of
codimension two is again a $\rho ^{-7/2}$ tail in the pdf of the density.  

For $d=3$ another type of singularity with vanishing Hessian has been
identified, denoted by $A_1A_3$ in Ref.~\cite{boga}. It corresponds to a
kurtoparabolic point at which the tangent (hyper)plane has another point of
tangency with the graph of the Lagrangian potential. Such points do not
contribute to the leading order of the pdf for large densities. For $d>3$,
higher-order singularities such as the $A_5$-points of Ref.~\cite{boga} can
appear generically and we do not know how they affect the -7/2 law.

\section{Concluding remarks}
\label{s:conclusion}

In this paper we determined the pdf of the mass density for the
Burgers equation in the inviscid limit for smooth initial conditions
which are random but not necessarily homogeneous. We showed that in
one, two and three dimensions this pdf has a power-law tail
with exponent \hbox{-7/2}.  In one dimension this tail originates from
preshocks, that is nascent shocks which take place at discrete times,
whereas in two and three dimensions the tail comes from
time-persistent boundaries of shocks (associated in
Lagrangian space to kurtoparabolic singularities).  In the
neighborhood of such points, arbitrary large but finite densities are
present.  Note that, in any case, the -7/2 law is due to a phenomenon
of shock germination, either in space-time or just in space.

The density pdf will of course be modified if, instead of
the limit of vanishing viscosity, we assume a finite but small
viscosity. A similar question has already been considered by Gotoh and
Kraichnan \cite{GotohKraichnan} for the pdf of large (negative)
velocity gradients $\xi$ in one-dimensional forced Burgers turbulence;
they found that, at very large values of $\xi$ there is a power-law
range with exponent -1, different from the exponent prevailing at
those values of $\xi$ where the inviscid limit is achieved.
Similarly, we expect that, due to the large shear inside shocks, the
-7/2 tail in the density pdf would become just an intermediate
asymptotic range, beyond which another law should prevail.

\begin{figure}
\iffigs 
\centerline{\psfig{file=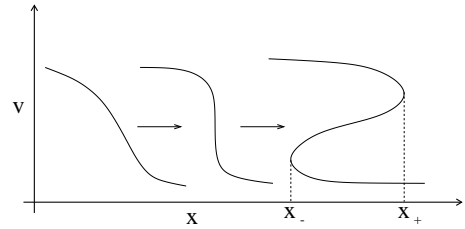,width=6cm}}
\else\drawing 65 10 {Normal form prechoc 1D}
\fi
\vspace{2mm}
\caption{Position-velocity phase space for the
(Jeans)--Vlasov--Poisson equation in one dimension. Support of the
particle distribution at various times for
single-speed initial conditions.}
\label{f:vlasov}
\end{figure}
As mentioned in the Introduction, the Burgers equation with vanishing
viscosity is used by cosmologists, under the name of adhesion model,
to approach the problem of the formation of large scale structures.
In principle, the appropriate mathematical framework should involve
partial differential equations for density and velocity, coupled to
the Poisson equation for the gravitational potential. The initial
velocity (in an expanding universe at decoupling) is then determined
uniquely in terms of initial density fluctuations (see, e.g.,
Section~2.2.2 of Ref.~\cite{VergassolaDubrulleFrischNoullez}). When
dealing with dustlike {\em collisionless\/} matter, a hydrodynamical
description cannot be justified on the usual grounds that local
thermodynamic equilibrium is quickly achieved.  However, as long as
particles do not cross, a quasi-hydrodynamical description, without
any viscous diffusion term, is suitable. After crossing, the
multi-stream situation may be described pseudo-microscopically
in terms of a distribution function $f(\x,\v,t)$ in the
position-velocity phase space satisfying the $d$-dimensional
(Jeans)--Vlasov--Poisson \cite{henon}
\begin{equation}
\partial_tf(\x,\v,t) +\left(\v\cdot
\nabla_{\x}-\nabla_{\x}\cdot\nabla_{\v}\right) f=0,
\label{jvp}
\end{equation}
supplemented by the Poisson equation, relating the gravitational
potential to the density $\rho(\x,t)\equiv \int f(\x,\v,t)\,d\v$. (We
omitted the expansion factor for simplicity.)  In (\ref{jvp}), the
position and velocity variables are in principle independent, but the
relevant solutions for the pseudo-microscopic description are of the
``single-speed'' type. By this we understand that the distribution $f$
has its support on a $d$-dimensional submanifold of the
$2d$-dimensional phase space such that, initially, there is a single
velocity $\u_0(\x)$ associated to a given position $\x$.  Such
single-speed solutions may, after particle crossing, possess more than
one velocity for a given $\x$.  The distribution $f$ does however
remain well-defined (see Fig.~\ref{f:vlasov}, which corresponds to the
one-dimensional case).  For smooth initial data, the support of $f$
remains smooth even after particle crossing (see Ref.~\cite{roytvarf}
for the one-dimensional case). In one dimension, it is easy to show
that, near a point $x_*$ where the tangent to the graph of the support
is parallel to the $v$-axis (such as points $x_\pm$ on
Fig.~\ref{f:vlasov}), the density has a singularity $\propto
|x-x_*|^{-1/2}$ \cite{Arnold}.  When $x_*$ is random with a
probability density at $x_*=x$, we infer by an argument similar to
that of Section~\ref{s:7halvlaw} that the pdf $p(\rho;x,t)\propto \rho
^{-3}$ for $\rho\to\infty$. This argument carries over to higher 
dimensions; the  $x_*$-points are then on the $(d-1)$-dimensional
manifold where the Jacobian of the Lagrangian map vanishes.

The same $\rho ^{-3}$ law holds within the $d$-dimensional Zel'dovich
approximation \cite{kofman}. The latter is indeed equivalent to a modified
prescription for determining the gravitational potential; it is
obtained from the Poisson equation only initially and its gradient
(the gravitational force) is then taken constant along particle
trajectories.  Clearly, the modified prescription has the only effect
that it changes the precise position of the support (after the first
crossing) but not the nature of the ensuing density singularities. As
noticed in Ref.~\cite{EVriemann} the power law pdf with exponent $-3$
relative to the velocity gradient for the inviscid one-dimensional
forced Burgers equation, proposed in Ref.~\cite{GotohKraichnan}, can
be interpreted in terms of singularities if one allows multi-valued
solutions; this is precisely the case in collisionless physical
situations.

The Burgers/adhesion model has been found useful for describing large-scale
features of collisionless dynamics, such as the positions and slow thickening
of pancakes. It is nevertheless intrinsically a {\em hydrodynamical\/} model
and, as such, better suited to handle low-pressure collision-dominated
matter. As far as their singularities are concerned, the Zel'dovich
approximation and the Burgers/adhesion model are in different universality
classes and have different tail behavior for the density pdf. We finally
mention that with the three-dimensional adhesion model, various other
quantities may be calculated analytically, which are more directly related to
what can be determined by cosmological observations and/or by N-body
simulations, such as the density correlation function (at small distances) and
the pdf of the mass in a ball of small radius.

\vspace*{5mm}
\par\noindent {\bf Acknowledgements}
\vspace{3mm}

We are grateful to V.I.~Arnold for pointing out to UF in 1992 the full 2-D
singular structure, to I.A.~Bogaevski for discussions on convex hulls of 3-D
surfaces embedded in 4-D space, to S.~Gurbatov for drawing our attention to
eq.~(\ref{gurbrel}) and to J.~Farley for help with Pascoli's poem. Part of
this work was done while JB and UF where visiting the Center of Nonlinear
Studies (Los Alamos, USA) and the Isaac Newton Institute (Cambridge, UK);
their support is gratefully acknowledged. This work was also supported by the
European Union under contract HPRN-CT-2000-00162.


\end{document}